\documentclass[a4paper,aps,pre,superscriptaddress,floatfix,nofootinbib,twocolumn,showpacs]{revtex4-1}

\usepackage{amsmath}
\usepackage{amsfonts}
\usepackage{amssymb}
\usepackage{graphicx}
\usepackage[usenames]{color}
\usepackage[latin1]{inputenc}
\usepackage{float}
\usepackage{comment}
\usepackage{grffile}
\usepackage{hyperref}

\newcommand{\av}[1]{\langle #1 \rangle}
\newcommand{\fluc}[1]{\langle {#1}^2\rangle}
\newcommand{\avk}{\langle k \rangle}
\newcommand{\fluck}{\langle k^2\rangle}
\newcommand{\be}{\begin{equation}}
\newcommand{\ee}{\end{equation}}
\newcommand{\bea}{\begin{eqnarray}}
\newcommand{\eea}{\end{eqnarray}}
\newcommand{\km}{k_{\rm{max}}}
\newcommand{\dt}{\Delta t}

\begin{document}
\title{Epidemic thresholds of the Susceptible-Infected-Susceptible
  model on networks: A comparison of numerical and theoretical
  results}
 
\begin{abstract}
Recent work has shown that different theoretical approaches
to the dynamics of the Susceptible-Infected-Susceptible (SIS) model for 
epidemics lead to qualitatively different estimates for the
position of the epidemic threshold in networks.
Here we present large-scale numerical simulations of the SIS
dynamics on various types of networks, allowing the precise determination
of the effective threshold for systems of finite size $N$. We compare
quantitatively the numerical thresholds with theoretical predictions
of the heterogeneous mean-field theory and of the quenched mean-field
theory. We show that the latter is in general more accurate, scaling
with $N$ with the correct exponent, but often failing to capture
the correct prefactor.

\end{abstract}

\author{Silvio C. Ferreira} 
\affiliation{Departamento de F\'{\i}sica, Universidade Federal de
  Vi\c{c}osa, 36571-000, Vi\c{c}osa - MG, Brazil}

\author{Claudio Castellano} 
\affiliation{Istituto dei Sistemi Complessi (ISC-CNR), via dei Taurini 19,
I-00185 Roma, Italy}
\affiliation{Dipartimento di Fisica, ``Sapienza''
  Universit\`a di Roma, P.le A. Moro 2, I-00185 Roma, Italy}

\author{Romualdo Pastor-Satorras} 
\affiliation{Departament de
  F\'{\i}sica i Enginyeria Nuclear, Universitat Polit\`ecnica de
  Catalunya, Campus Nord B4, 08034 Barcelona, Spain}

\maketitle

\section{Introduction}
\label{sec:intro}

Models for disease propagation are a paradigmatic example of processes
for which the interplay of a simple dynamics and a topologically
complex interaction pattern~\cite{barabasi02,Dorogovtsev:2002} gives
rise to nontrivial
phenomena~\cite{barratbook,dorogovtsev07:_critic_phenom,
  Newman10,Vespignani12}.  In this context, the
Susceptible-Infected-Susceptible (SIS) model \cite{anderson92} plays a
paramount role, as the simplest model undergoing an epidemic phase
transition between an absorbing \cite{Marrobook}, healthy phase, where
infection rapidly disappears, and an active phase, with a stationary
endemic state characterized by a finite fraction of infected
individuals.  In the SIS model nodes can be either susceptible or
infected. An infected node spreads the disease to each one of its
susceptible contacts at rate $\lambda$, while it heals at a rate
$\mu$, usually taken to be 1. A critical value $\lambda_c$ of the
spreading rate separates the absorbing phase ($\lambda \le \lambda_c$)
from the endemic one ($\lambda>\lambda_c$).

Traditional mathematical epidemiology has analyzed the behavior of the
SIS model mainly within a mean-field (homogeneous mixing)
approximation~\cite{anderson92}. In this case, the epidemic transition
occurs for a finite value $\lambda_c$ of the control parameter,
inversely proportional to the average connectivity $\avk$ of the
interaction pattern.  This view was completely revolutionized
after the study of SIS on networks with large variations of nodes'
connectivities. The first mathematical
approaches to SIS in networks \cite{Pastor01b, Romu} were based on
the heterogeneous mean-field (HMF) approach
\cite{dorogovtsev07:_critic_phenom}, an extension of standard
mean-field theory taking explicitly into account the large
fluctuations of the degree $k$ (number of connections) of single
nodes, and that is based on replacing the actual topological
structure of the network, given by its adjacency matrix
$A_{ij}$\footnote{A matrix taking value $1$ if
nodes $i$ and $j$ are connected and zero otherwise.},
by an average ${\bar A}_{k_i,k_j}$, expressing the probability
that two vertices of degree $k_i$ and $k_j$ are connected
in the original network (the so-called \textit{annealed network
  approximation} \cite{dorogovtsev07:_critic_phenom}).

From here, an
expression for the epidemic threshold is derived for uncorrelated
networks\footnote{Networks such that the probability that an edge
  arrives at a node of degree $k$ is proportional to $k P(k)$
  \cite{Dorogovtsev:2002}.} \cite{boguna2002} 
\begin{equation}
  \lambda_c^{\mathrm{HMF}} =  \frac{\avk}{\fluck},
  \label{lambda_HMF}
\end{equation}
where $\av{k^n} = \sum_k k^n P(k)$ is the $n$-th moment of the
network's degree distribution $P(k)$
\cite{barabasi02}. Eq.~(\ref{lambda_HMF}) predicts the epidemic
threshold to vanish for networks with diverging second moment
(``scale-free''), and to be finite otherwise.  This result has huge
implications, since many real networks have a power-law degree
distribution $P(k) \sim k^{-\gamma}$ with $\gamma<3$, so that $\fluck$
grows unboundedly with the system size
\cite{barabasi02,Dorogovtsev:2002,Newman10}.

For almost a decade, HMF results for the SIS model were considered
essentially correct in the case of random networks within the
statistical physics community\footnote{Its validity in random regular
  graphs has been however contested by means of a HMF pair
  approximation theory \cite{Eames01102002,Gleeson11}}, although no
systematic detailed investigation of their accuracy was
performed~\cite{Hwang05}. In other communities, however, different
theoretical approaches have been applied, yielding opposite results.
Chatterjee and Durrett~\cite{Chatterjee09, Durrett10} have rigorously
proven that, for strictly infinite system size, the epidemic threshold
is exactly $0$ for any exponent $\gamma$ of the degree distribution.
Although of fundamental importance, this result does not provide a
simple understanding of the physical origin of the threshold
vanishing.  Within the computer science community, another approximate
approach, that we term quenched mean-field (QMF) theory, has been put
forward by Wang et al.~\cite{Wang03,1284681}.  The basic idea is to
write down the evolution equation for the probability $\rho_i$ that a
certain node $i$ is infected.  Taking into account the actual
connections in the network, as given by the adjacency matrix, this
approach predicts the existence of a threshold given, for any graph,
by the inverse of the largest eigenvalue of the adjacency matrix,
namely
\begin{equation}
  \lambda_c^{\mathrm{QMF}} = \frac{1}{\Lambda_N}.
  \label{QMF}
\end{equation}
This result has been obtained later also by Gomez et al.~\cite{Gomez10}
and, in a more refined way, by Van Mieghem et al.~\cite{VanMieghem09}.

Equation~(\ref{QMF}) can be complemented with the expression for
$\Lambda_N$ obtained by Chung et al.~\cite{Chung03}, to obtain an
explicit estimate of the epidemic threshold~\cite{Castellano10}:
\begin{equation}
  \lambda_c^{\mathrm{CL}} = 
  \left \{ 
    \begin{array}{lr}
      c_1 /\sqrt{\km} &~~~~~~~ \sqrt{\km} > \frac{\fluck}{\avk}
      \ln^2(N) \\ 
      c_2  /\frac{\fluck}{\avk} & ~~~~~~~\frac{\fluck}{\avk} >
      \sqrt{\km} \ln(N) 
    \end{array}
  \right. ,
\label{lambda_chung}
\end{equation}
where $\km$ is the largest degree in the network and $c_1$ and $c_2$
are numerical constants.  This formula agrees with the result of
Ref.~\cite{Chatterjee09} since the threshold vanishes for large $N$ as
soon as $\km$ diverges with $N$, independently of 
$\gamma$ \cite{Dorogovtsev:2002}.
The threshold~(\ref{lambda_chung})
scales as the HMF result only for $\gamma < 5/2$.  The validity of the
predictions of Eq.~(\ref{lambda_chung}) has been qualitatively
verified for scale-free and scale-rich networks in
Ref.~\cite{Castellano10}.  The physical origin of the different
scaling for $\gamma$ larger or smaller than $5/2$ has been clarified
in Ref.~\cite{Castellano12}.  However, a detailed investigation of the
accuracy of the different theoretical approaches for generic graphs is
still lacking.

It is important to stress that Eq.~(\ref{QMF}) is an improvement over
HMF, but it is not exact.  While in HMF theory the actual network
structure is replaced by an annealed one \cite{Boguna09}, QMF theory
fully preserves the detailed quenched structure of the network as
described by the adjacency matrix. In fact, Eq.~(\ref{lambda_HMF}) can
be derived from Eq.~(\ref{QMF}) by inserting the largest eigenvalue of
the annealed network\footnote{In annealed networks, the adjacency matrix
  takes a probabilistic interpretation, being replaced by the
  probability that two vertices $i$ and $j$ are connected. In degree
  uncorrelated networks, this annealed network adjacency matrix takes
  the form \cite{dorogovtsev07:_critic_phenom}
  \begin{equation}
    \label{eq:1}
    \bar{A}_{k_i,k_j} = \frac{k_i k_j}{\av{k}N}.
  \end{equation}
  From this expression we can easily see that $v_i = k_i$ is an
  eigenvector of $\bar{A}$ with eigenvalue $\fluck/\avk$. We can also
  see that $\mathrm{Tr}(\bar{A}) = \fluck/\avk$.  Since $\bar{A}$ is a
  positive semidefinite matrix, all its eigenvalues are non negative,
  and since the trace is equal to the sum of the eigenvalues, it
  follows that the largest eigenvalue of the adjacency matrix in
  annealed networks is equal to $\fluck/\avk$, all the rest of the
  eigenvalues being equal to zero.}.  In this sense the HMF approach
is equivalent to QMF theory plus an additional, annealed network,
approximation.  Yet both approaches rely on the mean-field
(single-particle) assumption that the probability that nearest
neighbor nodes are active can be factorized as the product of the
single node probabilities. They thus neglect possibly important
dynamical correlations between the state of adjacent nodes.  For this
reason, while it is appropriate to say that QMF approach improves over
HMF, there is no guarantee about the exactness of Eq.~(\ref{QMF}),
whose accuracy needs to be checked on a case-by-case basis.

In this paper we tackle this task, by performing large-scale numerical
simulations of the SIS model on various types of graph, using the
quasi-stationary method to study the phase-transition in finite
systems \cite{DeOliveira05}, and determining the effective
size-dependent threshold by analyzing the peak of the susceptibility
\cite{Binder2010}. The accuracy of the method is checked by applying
it on heterogeneous annealed networks
\cite{gil05:_optim_disor,stauffer_annealed2005,Boguna09}, in which HMF
is exact \cite{Boguna09,FFPS11}. We then consider an example of homogeneous
network, the random regular graph, and an instance of strongly
heterogeneous network, the star network.  In the first case, we
observe that both HMF and QMF predict finite thresholds, close to the
numerical one, but not equal to it. In the second case, QMF correctly
predicts the scaling of the vanishing threshold with network size, but
the prefactor is not the right one. Finally, we consider the key case
of power-law degree distributed networks.  For $\gamma<5/2$ both
theories turn out to give an asymptotically exact value for the
threshold. For $5/2 < \gamma <3$ we observe a vanishing threshold
whose scaling with $N$ is correctly predicted by QMF, although with an
incorrect prefactor. For $\gamma>3$ we numerically recover the
presence of two competing activation mechanisms discussed in
Ref.~\cite{Castellano12}: QMF prediction follows approximately the
transition due to the activation of the largest hub in the network,
which vanishes and dominates in the large size limit. The HMF
threshold remains finite and is close to the transition point due to
the most densely connected core of the network (maximum  $k$-core).

Our numerical results confirm that QMF is indeed an improvement over
HMF, providing a better estimate of the threshold and capturing the
vanishing threshold for power-law distributed networks with any
$\gamma$, a key fact missed by the HMF approach. However, it is only
the scaling of the threshold with network size that is correctly given
by QMF. Improved analytical strategies, beyond quenched mean-field are
thus needed in order to obtain more accurate threshold predictions in
cases of practical importance. After submission of this paper, we
became aware of a publication \cite{PhysRevE.86.026116} in which a
similar numerical comparison is performed on much smaller network
sizes than those considered here. 

\section{Numerical methods}
\label{sec:peak}

We consider here the SIS model for epidemics in continuous time. The
numerical algorithm is implemented as follows: At each time step, we
compute the number of infected nodes, $N_i$, and links emanating from
them, $N_n$.  With probability $N_i/(N_i+\lambda N_n)$ one infected
node, chosen at random, becomes healthy. With complementary
probability $\lambda N_n/(N_i+\lambda N_n)$, one of the links is
selected uniformly at random and the infection is transmitted through
it from the infected node corresponding to one of the ends of the
edge, towards the (possibly susceptible) node at the other end. The
numbers of infected nodes and related links are updated accordingly,
time is incremented by $\dt = 1/(N_i+\lambda N_n)$, and the whole
process is iterated.

\subsection{The quasi-stationary state method}
\label{sec:quasi-stationary-}

The standard numerical procedure to investigate the properties of
absorbing phase transitions is based on the determination of the
average of the order parameter (in this case the density of infected
nodes), restricted only to surviving runs \cite{Marrobook}, i.e., runs
which have not reached the absorbing state up to a given time
$t$. Such a technique is quite wasteful, because surviving
configurations are very rare at long times close to the threshold, and
an exceedingly large number of realizations of the process are needed
in order to get substantial statistics. An alternative technique is
the quasi-stationary state (QS) method~\cite{DeOliveira05,FFPS11},
based on the idea of constraining the system in an active state.  This
procedure is implemented by replacing the absorbing state, every time
the system tries to visit it, with an active configuration randomly
taken from the history of the simulation. For this task, a list of $M$
active configurations, corresponding to states previously visited
  by the dynamics, is stored and constantly updated.  An update
consists in randomly choosing a configuration in the list and
replacing it by the present active configuration with a probability
$p_{r}\Delta t$.  After a relaxation time $t_r$, the QS quantities are
determined during an averaging time $t_a$. The QS probability
$\bar{P}_n$ that $n$ vertices are infected is computed during the
averaging interval, each configuration with $n$ active vertices
contributing to the QS distribution with a probability proportional to
its lifespan $\Delta t$. From the particle distribution $\bar{P}_n$,
the moments of the activity distribution can be computed as
\begin{equation}
  \label{eq:8}
  \langle \rho_s^k \rangle = \frac{1}{N}\sum_{n\ge1} n^k
\bar{P}_n. 
\end{equation}
The simulation procedure described above was recently used to
determine with high numerical accuracy the critical point and
exponents of the contact process~\cite{Marrobook} on
annealed~\cite{FFPS11} and quenched~\cite{FFCR11} networks with power
law degree distributions.  

The values of the QS parameters used in the present simulations were
$M=300$ and $p_r= 0.02$, while $t_r$ and $t_a$ varied depending on $N$
and $\lambda$.

\subsection{Numerical determination of the epidemic threshold}
\label{sec:numer-determ-epid}

In usual absorbing-state phase transitions, where the critical point
converges to a finite value in the thermodynamic limit, the
finite-size scaling method allows the precise numerical determination
of the critical point and the whole set of associated critical
exponents~\cite{Marrobook}.  It is based on the computation, for
increasing system sizes $N$, of $\rho_s$, the global activity density,
computed from surviving run averages or from QS simulations.  For
values of the control parameter $\lambda$ in the active phase,
$\rho_s$ reaches a finite non-zero limit as $N \to \infty$. For values
of the control parameter in the absorbing phase, $\rho_s$ decays
trivially as $\rho_s \sim N^{-1}$, since there is essentially of the
order of one active vertex in the whole network. The critical point is
distinguished by a power-law decay $\rho_s \sim N^{-\beta/\bar{\nu}}$,
where $\beta$ is the critical exponent controlling the density
$\rho_s$ at a finite distance from the critical point, while
$\bar{\nu}$ is associated to the growth of correlations close to
criticality \cite{Marrobook}.

In networks with a diverging second moment for the degree
distribution, this approach can go awry if strong corrections to
scaling are present \cite{FFPS11,FFCR11}. In the particular case of
the SIS model on networks with a power-law degree distribution,
this approach simply does not work, because the effective threshold
depends on $N$ and it goes to zero as the system size
grows. Therefore, for any value of $\lambda>0$, $\rho_s$ will attain a
finite limit for a sufficiently large $N$, once the corresponding
$\lambda_c(N)$ becomes smaller than $\lambda$. This shows that
asymptotically the threshold is zero~\cite{Castellano10}, but it does
not provide information on the effective threshold for finite $N$.  To
overcome this problem, we turn instead to another procedure to
determine the critical point, namely the study of the susceptibility
\cite{Binder2010}, defined as
\begin{equation}
  \label{eq:9}
  \chi = N \frac{\fluc{\rho}-\av{\rho}^2}{\av{\rho}}.
\end{equation}
When plotted as a function of $\lambda$ in a system of fixed size $N$,
the susceptibility $\chi$ exhibits a maximum at a value
$\lambda_p(N)$. In systems with a finite critical point in the
thermodynamic limit $\lambda_c(\infty)$, the peak of the
susceptibility at $\lambda_p(N)$ corresponds to a transition rounded
by finite size effects, that as $N \to \infty$ tends to the critical
point as $\lambda_p(N) - \lambda_c(\infty) \sim
N^{-1/\bar{\nu}}$~\cite{Binder2010}.  Correspondingly, the height of
the susceptibility at the maximum scales with system size as
$\chi^\mathrm{max} \sim N^{\gamma'/\bar{\nu}}$, where $\gamma'$ is a
new critical exponent.

We assume that the relation between $\lambda_p$ and $\lambda_c$ written
above holds also in the present case, where $\lambda_c$ depends on $N$:
$\lambda_p(N) - \lambda_c(N) \sim N^{-1/\bar{\nu}}$.
This implies that the susceptibility peak and the size-dependent threshold
tend to coincide in the large size limit. 
When the assumption can be explicitly controlled, it turns out
(see below) to be correct.

It is worth noticing that the susceptibility Eq.~\eqref{eq:9} is different from the
standard definition $\chi_N = N(\fluc{\rho}-\av{\rho}^2)$
\cite{Marrobook}. We adopt Eq.~\eqref{eq:9} because it leads to clearer
numerical results (see Sec.~\ref{sec:annealed-scale-free}), while
preserving all the scaling properties of the usual definition.

\section{Numerical check of the susceptibility method: Annealed
  scale-free networks}
\label{sec:annealed-scale-free}

A natural benchmark to check the accuracy of the susceptibility peak
as a measure of the position of the critical point in the SIS model is
given by annealed networks
\cite{gil05:_optim_disor,stauffer_annealed2005,Boguna09}.  In annealed
networks, all edges are rewired (preserving the degree and the degree
correlations of the involved nodes) after each change of the state of
any vertex.  This procedure destroys all dynamical correlations, and
thus renders exact the prediction of HMF theory
\cite{dorogovtsev07:_critic_phenom,Boguna09}. From a practical point
of view, the regeneration of the whole network every time a
microscopic dynamic step is performed can be effectively achieved in
uncorrelated networks by selecting at random, every time that a
nearest neighbor of a vertex is needed, a vertex of degree $k'$, with
probability $k'P(k')/\av{k}$~\cite{Boguna09}.
\begin{figure}[t]
  \centering
  \includegraphics[width=8cm,clip=true]{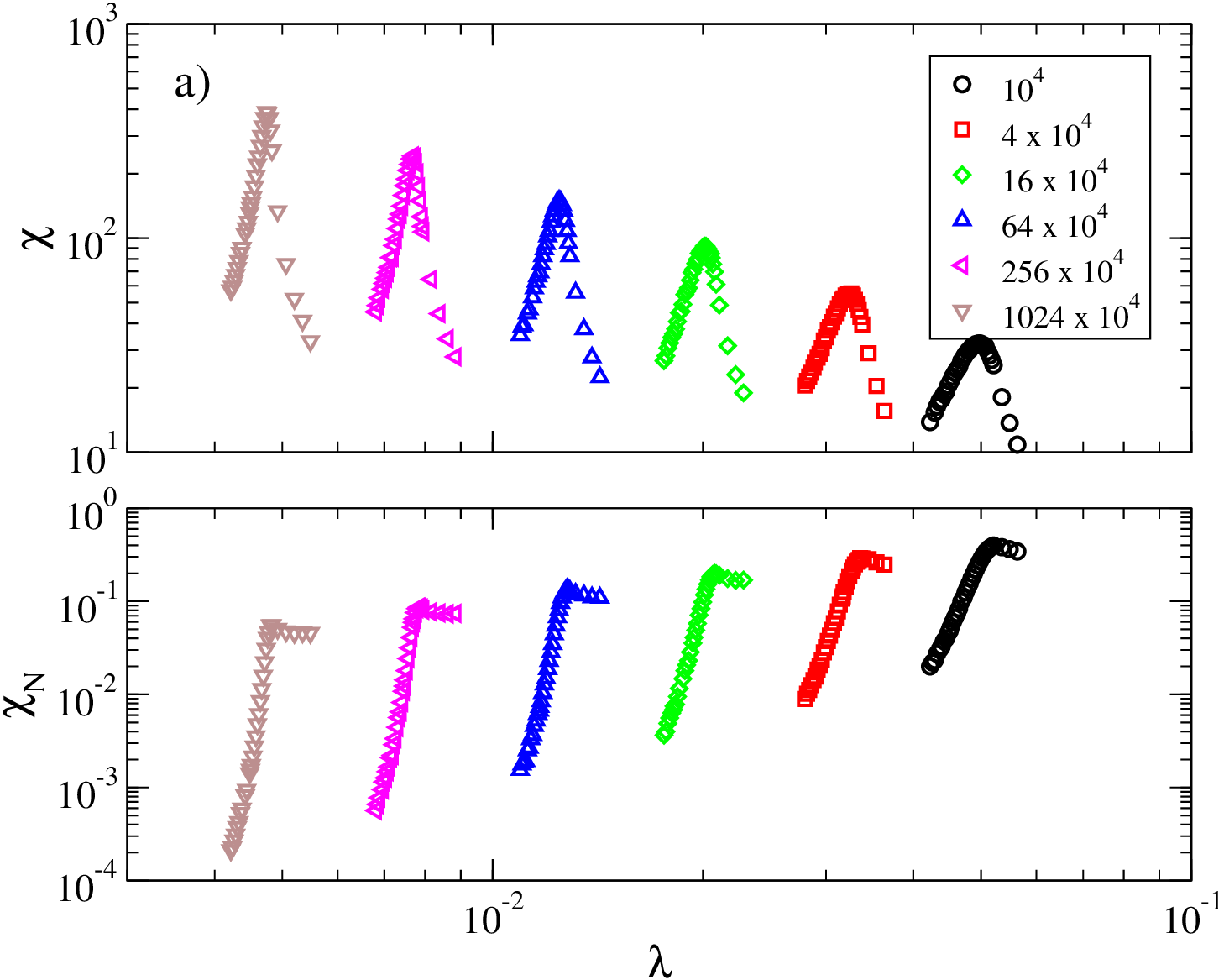}\\~\\
  \includegraphics[width=8cm,clip=true]{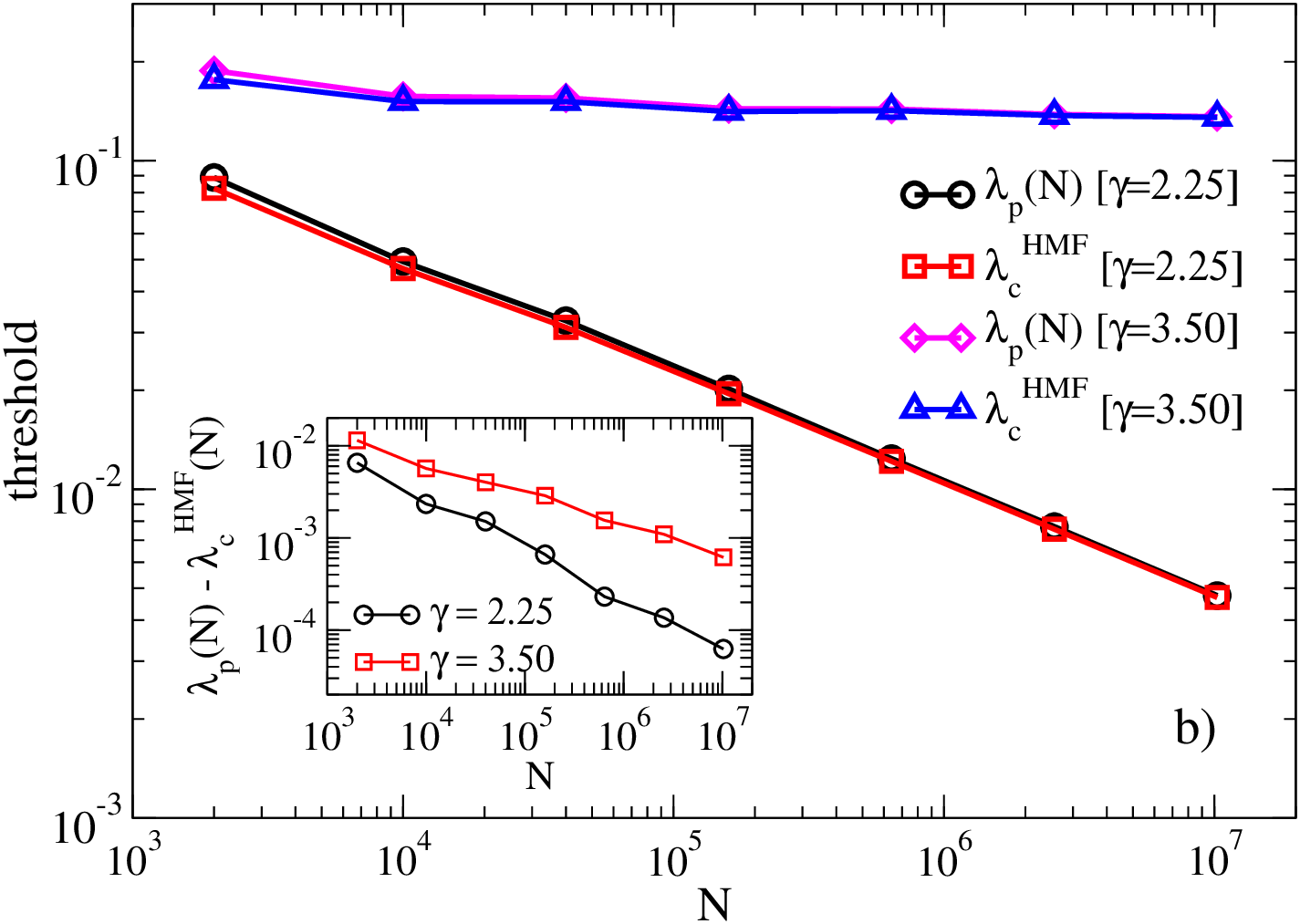}
  \caption{(Color online) (a) Plot of $\chi= N
    (\fluc{\rho}-\av{\rho}^2 )/\av{\rho}$ and $\chi_N= N
    (\fluc{\rho}-\av{\rho}^2)$ for the SIS process on annealed SF
    networks with degree exponent $\gamma=2.25$ and different network
    sizes. The susceptibility $\chi$ is more efficient to
    determine the effective size-dependent threshold. (b) Effective
    threshold from the susceptibility peak $\lambda_p(N)$ as a
    function of $N$, for annealed SF networks with $\gamma=2.25$ and
    $\gamma=3.5$, compared with HMF predictions. 
    The effective threshold shows a very good agreement
    with the numerically evaluated threshold $\lambda_c^\mathrm{HMF}$
    (Eq.~\eqref{lambda_HMF}). Inset: Differences between $\lambda_p(N)$
    and the theoretical predictions $\lambda_c^{HMF}(N)$ for $\gamma=2.25$
    and $\gamma=3.5$.}
  \label{comparison_annealed}
\end{figure}
In Fig.~\ref{comparison_annealed} we present the results of
susceptibility measurements for uncorrelated annealed networks with
degree exponent $\gamma=2.25$, and $3.5$.
Fig.~\ref{comparison_annealed}(a) depicts the susceptibility $\chi$
defined in Eq.~\eqref{eq:9} and the usual susceptibility $\chi_N$
defined in the analysis of absorbing phase
transition~\cite{Marrobook}, in networks with $\gamma=2.25$ and
different sizes $N$.  We observe that $\chi$ provides a more clear-cut
definition of the susceptibility peak. The same behavior is observed
for $\gamma=3.5$ (data not shown). In
Fig.~\ref{comparison_annealed}(b) we plot the evolution of the
susceptibility peak $\lambda_p(N)$ as a function of network size for
fixed $\gamma=2.25$ and $\gamma=3.5$, and compare it with the
numerically evaluated HMF prediction.  These results confirm that
$\lambda_p$ provides an excellent approximation of the exact result
$\lambda^{HMF}_c$, both when the threshold goes to zero with $N$ and
when it converges to a finite value. The differences observed might be
attributed to corrections to scaling, such as those presented in
systems with a finite threshold in the thermodynamic limit, see
Sec.~\ref{sec:numer-determ-epid}.  Therefore, in the rest of the paper
we use the position of the peak of the susceptibility $\chi$ as the
numerical estimate of the position of the threshold.

\section{Homogeneous networks: The random regular network}
\label{sec:homogeneous-networks}

As a first non-trivial application of the technique of the
susceptibility peak to evaluate the SIS epidemic threshold in
homogeneous networks, we consider the case of random regular networks
(RRN) that is, networks where all nodes have exactly the same degree
$k$, while links are randomly distributed among them, avoiding
self-connections and multiple connections. In this case, HMF theory
predicts trivially a constant threshold $\lambda_c^{HMF} = 1/k$. The
prediction of QMF theory takes exactly the same value, as can be
easily seen by applying Perron-Frobenius theorem~\footnote{ It
    is easy to see that $v_i=1$ is an eigenvector of the adjacency
    matrix $A$ with eigenvalue $k$. Therefore, the result
    $\Lambda_N=k$ derives directly from Perron-Frobenius
    theorem~\cite{gantmacher}.}.
\begin{figure}[t]
 \centering
 \includegraphics[width=8.2cm,clip=true]{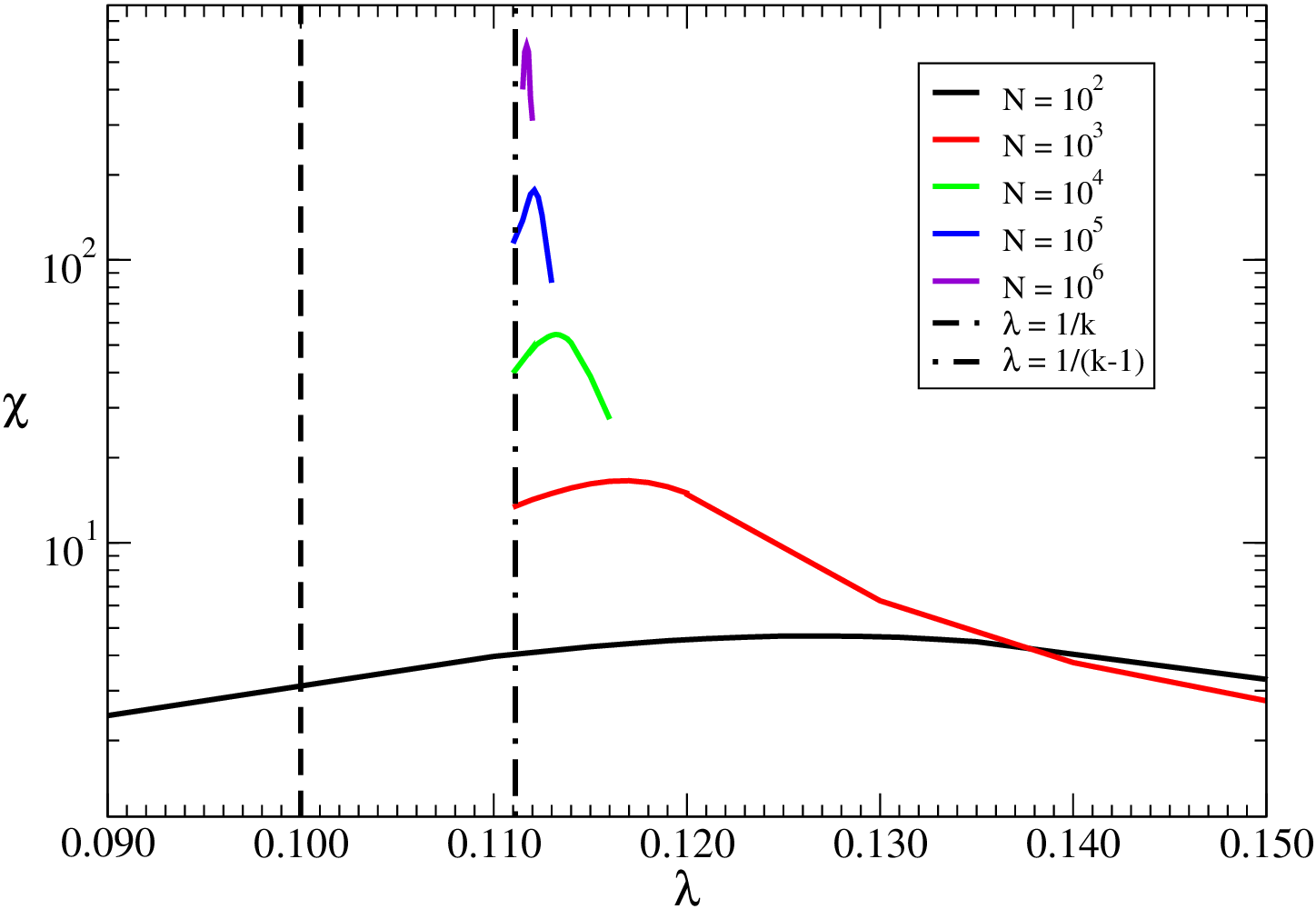}
 \caption{(Color online) Susceptibility as a function of $\lambda$ for
   RRN of increasing size (from bottom to top) and degree $k=10$. 
The susceptibility peak is closer to the theoretical
   prediction of the pair approximation than to the HMF and QMF
   results.}
\label{rrg}
\end{figure}
Fig.~\ref{rrg} shows the susceptibility $\chi$ as a function of
$\lambda$ for RRNs with increasing $N$ and degree $k=10$.  The
numerical estimated threshold is quite off from the theoretical value
$1/k$ for HMF and QMF, indicating that both theories are essentially
incorrect, while the susceptibility peak falls close (increasingly so
for larger $N$) to the value $\lambda_c^\mathrm{pair}=1/(k-1)$, which
is the prediction of the pair approximation \cite{Cator12}.

From this analysis we conclude that HMF and QMF provide a reasonable
approximation but not the exact position of the threshold.  They fail
just because they neglect dynamical correlations among the state of
neighbors, which are instead better taken into account by pair
approximation approaches \cite{Eames01102002,Gleeson11}.

\section{Heterogeneous networks: The star graph}
\label{sec:heter-netw}

In this Section we focus on the simplest case of a heterogeneous
network with vanishing epidemic threshold, namely the star network,
which is composed by a hub of degree $\km$, to which $\km$ leaves of
degree $1$ are attached.  For this star graph, the largest eigenvalue
of the adjacency matrix can be easily shown\footnote{The adjacency
  matrix of the star graph can be represented by $A_{1,j}=A_{j,1}=1$
  for $j\ge 2$ and $A_{i,j}=0$, otherwise. Therefore, if $v_i$ is an
  eigenvector of $A$, we have: (i) $\sum_{j=1}^ NA_{1,j}v_j =
  \sum_{j=2}^N v_j = \Lambda v_1$ and (ii) $\sum_{j=1}^ NA_{i,j}v_j =
  v_1 =\Lambda v_i$ for $i>1$. Replacing $v_i = v_1/\Lambda$ in
  equation (i) and using $N=\km+1$ we found the result $\Lambda =
  \sqrt{\km}$.  } to be $\sqrt{\km}$.  Therefore the QMF prediction
from Eq.~(\ref{QMF}) is $\lambda_c^{\mathrm{QMF}} = 1/\sqrt{\km}$. On
the other hand, the HMF result from Eq.~\eqref{lambda_HMF} takes in
this case the form $\lambda_c^{\mathrm{HMF}} = 2/(\km+1)$.
\begin{figure}[t]
 \centering
 \includegraphics[width=8.2cm,clip=true]{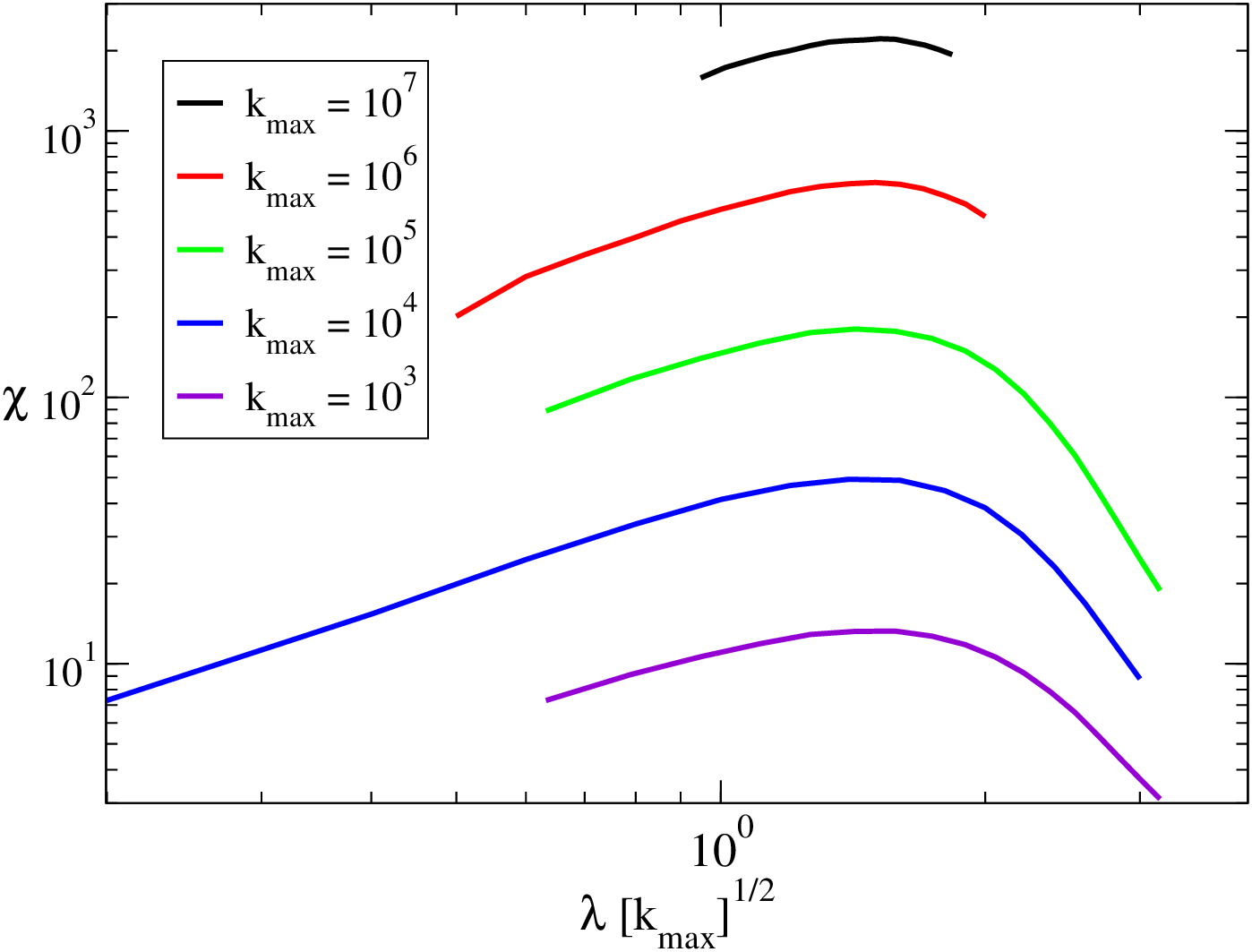}
 \caption{(Color online) Susceptibility $\chi$ as a function of
   $\lambda \sqrt{\km}$ computed in star graphs with different values
   of $\km$ (increasing from bottom to top).}
\label{stargraph}
\end{figure}
Figure ~\ref{stargraph} shows the susceptibility $\chi$ versus
$\lambda \sqrt{\km}$ computed for star graphs with a wide range of
values of $\km$. It clearly shows that the scaling $\lambda_c \sim
\sqrt{\km}$ is correct, however the value of the prefactor is around
$1.5$, rather than 1, in agreement with the rigorous bound $\lambda_c
\geq 1/\sqrt{\km}$ derived by Ganesh et al.~\cite{Ganesh05}.  The star
graph constitutes thus the simplest example of a network for which HMF
theory does not work.  This failure of HMF is altogether not
surprising, since this particular network is strongly correlated at
the degree level, and therefore fails to fulfill one necessary
condition for the validity of the HMF result Eq.~\eqref{lambda_HMF}.
QMF theory instead provides the correct scaling of the threshold with
network size, although the prefactor is not exact.

\section{Heterogeneous Networks: Power-law degree distributed graphs}
\label{sec:power-law-degree}

We now consider the SIS model on networks with power-law degree
distributions, $P(k) \sim k^{-\gamma}$, built using the uncorrelated
configuration model (UCM)~\cite{Catanzaro05}.  This procedure is equal
to the standard Molloy-Reed configuration model~\cite{Molloy95} with
the additional constraint that the degree values are strictly bounded
by $\km \sim N^{1/2}$. This bound guarantees that no topological
correlations are present in the network \cite{mariancutofss}, and
therefore fulfills the requirement needed for the applicability of the
HMF result Eq.~\eqref{lambda_HMF}.

We analyze three values of $\gamma$, representative of 
three regimes characterized by different expressions for the theoretical
estimates.

\subsection{$\gamma < 5/2$}

\begin{figure}[t]
 \centering
 \includegraphics[width=8.2cm,clip=true]{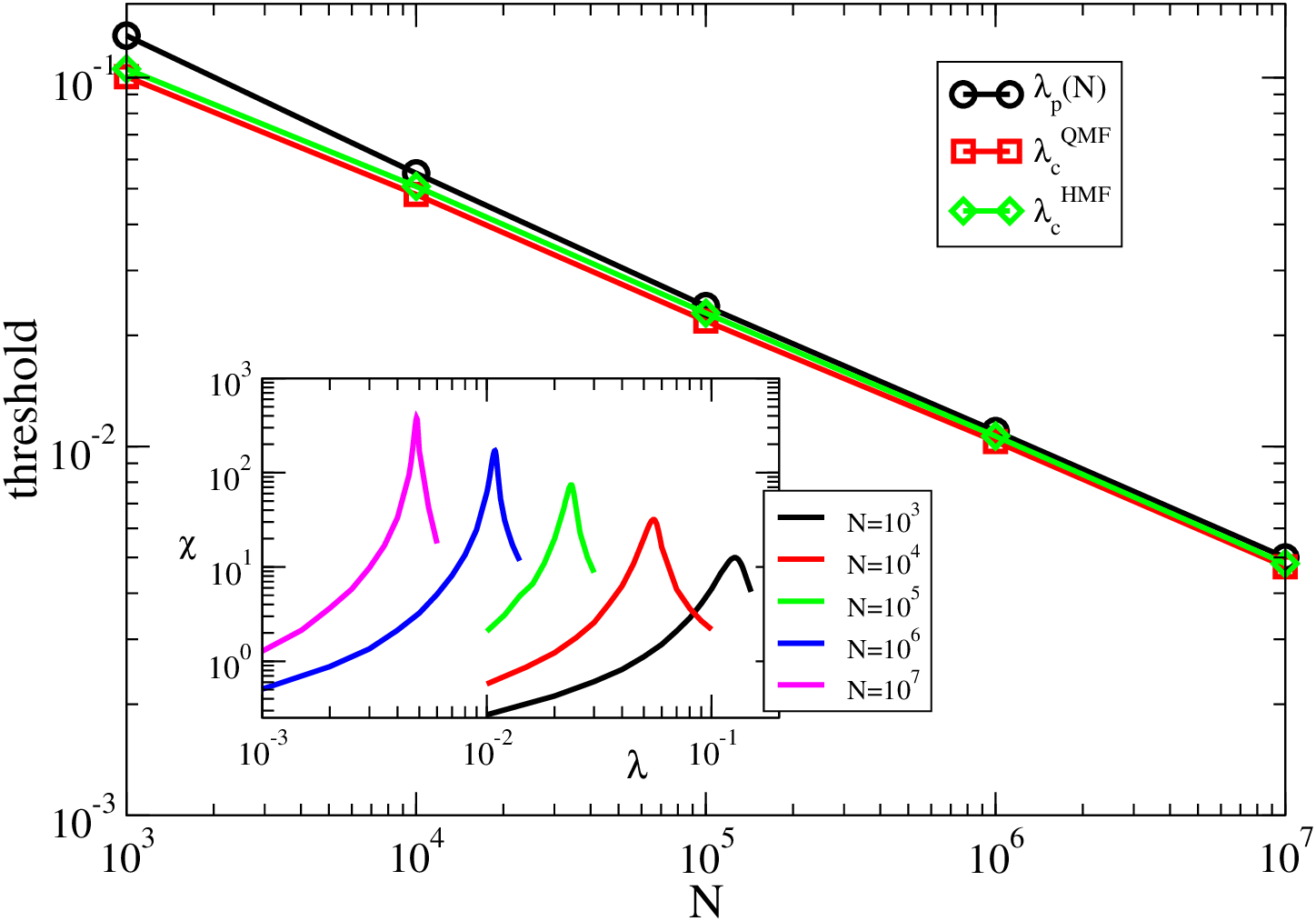}
 \caption{(Color online) Effective threshold $\lambda_p(N)$ from the
   susceptibility peak as a function of network size $N$ for
   uncorrelated SF networks with $\gamma=2.25$, compared with the HMF
   and QMF predictions. Inset: Susceptibility $\chi$ as a function of
   $\lambda$ for different network sizes (increasing from right
to left).} 
\label{g2.25}
\end{figure}

In Fig.~\ref{g2.25} we show the shape of the susceptibility $\chi$
versus $\lambda$ (inset) and the numerical threshold $\lambda_p$ as a
function of the network size $N$ (main plot) for $\gamma=2.25$,
compared with the predictions of the two theoretical approaches. It
turns out that the numerical results from the susceptibility peak
agree with good accuracy with both HMF and QMF theories, which in
their turn tend to coincide.  While it was expected that the
theoretical formulas scaled in the same way with $N$, see
Eq.~\eqref{lambda_chung}, the fact that they tend to coincide
indicates that the prefactor $c_2$ in Eq.~(\ref{lambda_chung}) is
$1$.  Fig.~\ref{g2.25} shows that HMF and QMF predictions are
apparently exact in the limit of large systems for $\gamma<5/2$.

In this particular range of $\gamma$, since the transition can
be identified with high accuracy, it is possible to extract additional
information about the epidemic phase transition. We consider thus the
scaling of the quantities $\rho_s$, $\chi$ and $\chi_N$ with $N$ at
the transition point. According to standard notation~\cite{Marrobook},
the expected scaling with system size should be (see
Sec.~\ref{sec:numer-determ-epid}): 
\be \rho_s \sim N^{-\beta/\bar{\nu}},
\;\;\chi_N \sim N^{\gamma'/\bar{\nu}}, \; \; \chi \sim
N^{(\gamma'+\beta)/\bar{\nu}}.
\label{eq:scaltheo}
\ee
\begin{figure}[t]
 \centering
 \includegraphics[width=8.2cm,clip=true]{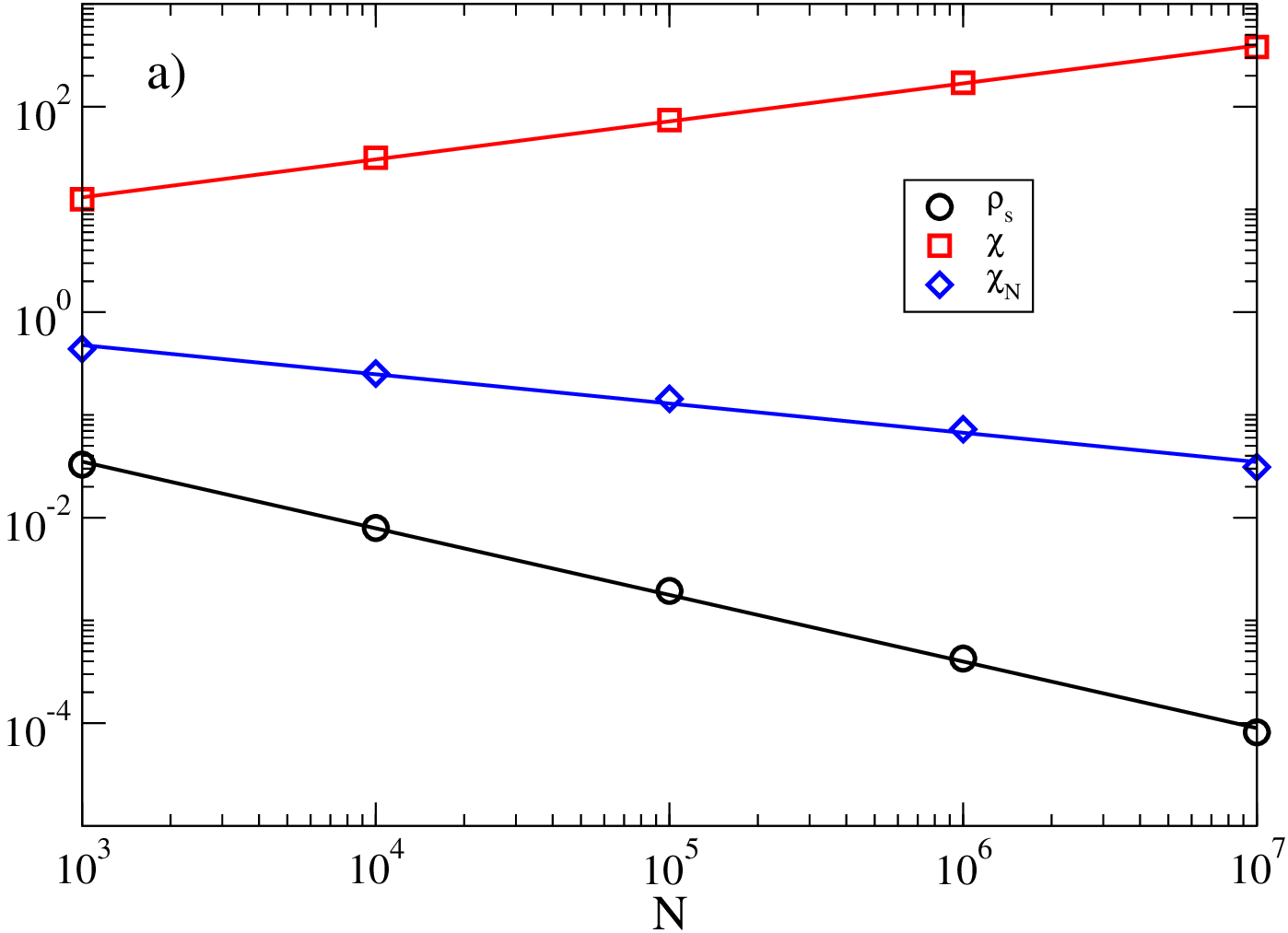}
 \\~\\
 \includegraphics[width=8.2cm,clip=true]{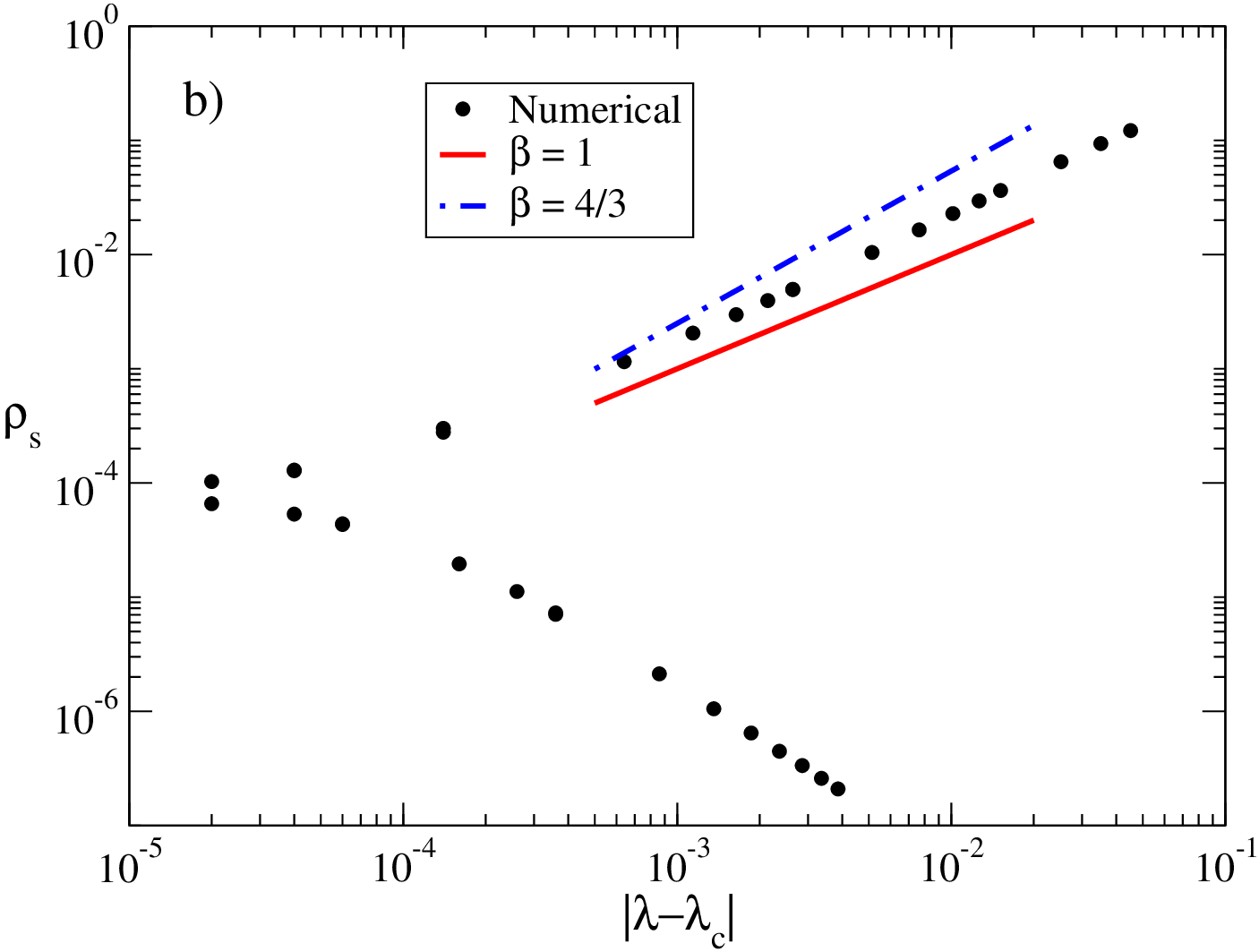}
 \caption{(Color online) (a) Numerical values of $\rho_s$, $\chi$ and
   $\chi_N$ as a function of $N$ evaluated at the susceptibility peak
   in SF networks with $\gamma=2.25$. (b) Stationary density of
   infected nodes as a function of the distance from the threshold
   $\lambda_p$ in SF networks with $\gamma=2.25$ and $N=10^7$. Lower
   points represent the subcritical phase.}
 \label{critscal}
\end{figure}
In Fig.~\ref{critscal}(a) we plot the values of $\rho_s$, $\chi$ and
$\chi_N$, evaluated at the susceptibility peak, as a function of $N$
in SF networks with $\gamma=2.25$. Fitting the curves in
Fig.~\ref{critscal}(a) to a power law form, we find the exponents: \be
\beta/\bar{\nu} = 0.65, \;\; \gamma'/\bar{\nu} = -0.28, \;\;
(\gamma'+\beta)/\bar{\nu} = 0.37.  \ee These exponents explain why
$\chi$ is the best choice to determine the threshold. The maximum of
the standard susceptibility $\chi_N$ scales with a negative exponent
$\gamma'$, and thus, in the limit of large $N$, the transition is
characterized by a discontinuity. The value $\gamma'+\beta>0$ instead
ensures a clearly defined maximum for the susceptibility $\chi$
diverging as $N\rightarrow\infty$.

By plotting the order parameter $\rho_s$ as a function of the
distance from the effective threshold, we can attempt to determine the
exponent $\beta$, which is defined by $\rho \sim
[\lambda-\lambda_c(N)]^\beta$. In Fig.~\ref{critscal}(b) we show such
a plot, for a SF network with $\gamma=2.25$ and size $N=10^7$.
According to HMF theory~\cite{Pastor01b}, the $\beta$ exponent is
expected to take the value $\beta=1/(3-\gamma)=4/3$, while the QMF approach
of Van Mieghem~\cite{VanMieghem12} predicts $\beta=1$.
The numerical results presented in Fig.~\ref{critscal}(b) yield an
effective exponent lying between the theoretical predictions,
so that the validity of none of them can be excluded.

\subsection{$5/2<\gamma<3$}

In this interval of $\gamma$, Eq.~(\ref{lambda_chung}) predicts that a different
regime sets in, with the threshold set by the inverse of the
square-root of $\km$ i.e. $\lambda_c^{\mathrm{CL}} \sim \km^{-1/2} =
N^{-1/4}$, while HMF theory predicts $\lambda_c^{\mathrm{HMF}} \sim
\km^{-(3-\gamma)} = N^{-(3-\gamma)/2}$ up to $\gamma=3$.  For the
values of $N$ which can be simulated numerically, the two theoretical
predictions are quite close but do not coincide.

\begin{figure}[t]
 \centering
 \includegraphics[width=8.2cm,clip=true]{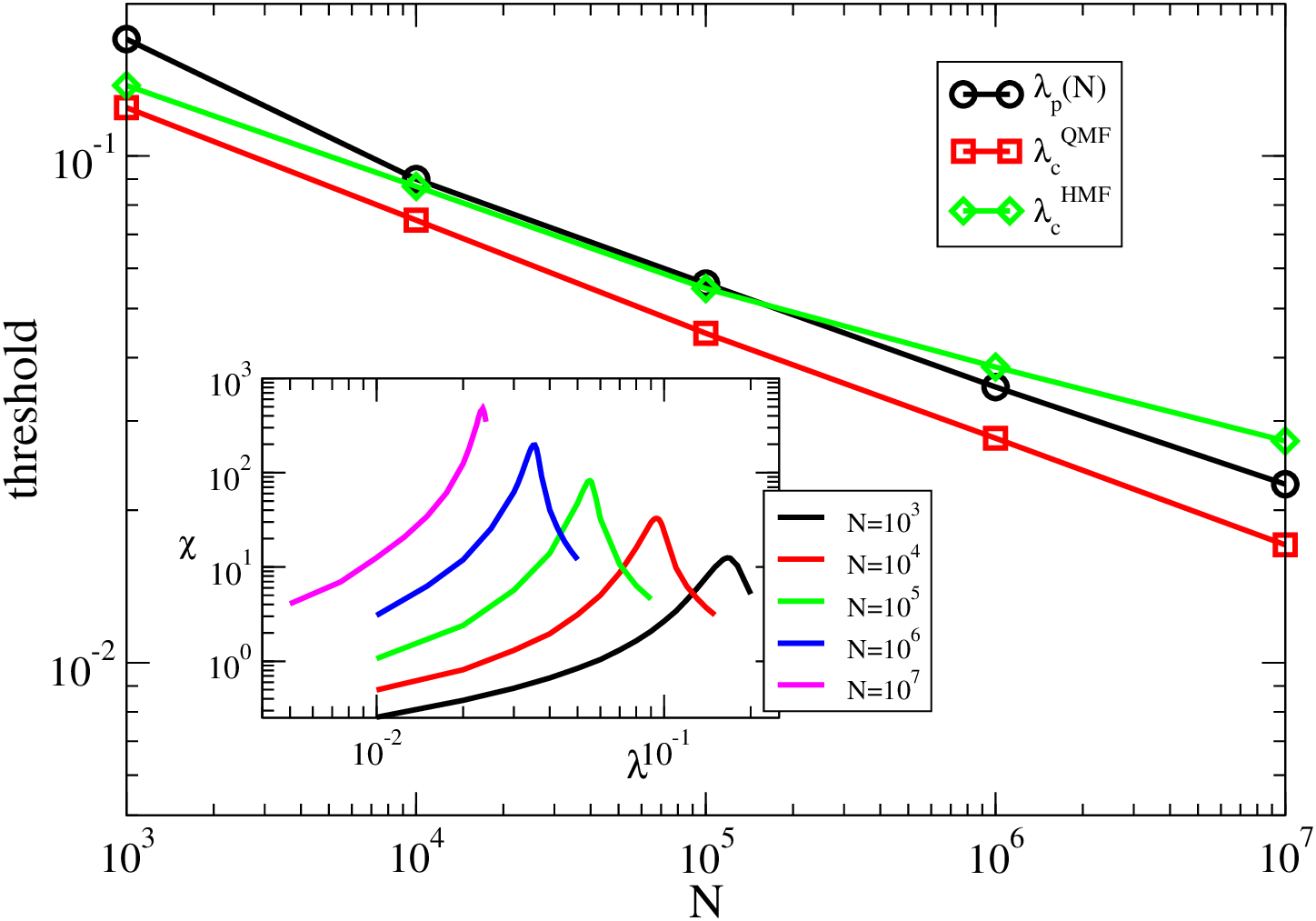}
 \caption{(Color online) Effective threshold $\lambda_p(N)$ 
as a function of network size $N$ for
   uncorrelated SF networks with $\gamma=2.75$, compared with the HMF
   and QMF predictions. Inset: Susceptibility $\chi$ as a function of
   $\lambda$ for different network sizes (increasing from right to
left).}
 \label{g2.75}
\end{figure}

Figure~\ref{g2.75} shows the results of the susceptibility analysis
for SF networks with $\gamma=2.75$. From this plot, we conclude that
the numerical results do not conform to the HMF behavior, the more so
for large system size. The numerical threshold $\lambda_p(N)$ scales
instead as the inverse of the largest eigenvalue, but with a
prefactor different from unity. The QMF threshold provides hence an approximation to the
numerical threshold, scaling in the same way, but with an accuracy of
the order of $30\%$.

\subsection{$\gamma > 3$}

For $\gamma>3$, HMF theory yields a finite value of the threshold,
which instead still vanishes according to QMF.  Since
sample-to-sample fluctuations of the value of $\km$ are quite large in this regime,
we consider for each value of $N$ only networks with $\km$ equal to
the mean value $\av{\km}$ \cite{Castellano10}.  

\begin{figure}[t]
 \centering
 \includegraphics[width=8.2cm,clip=true]{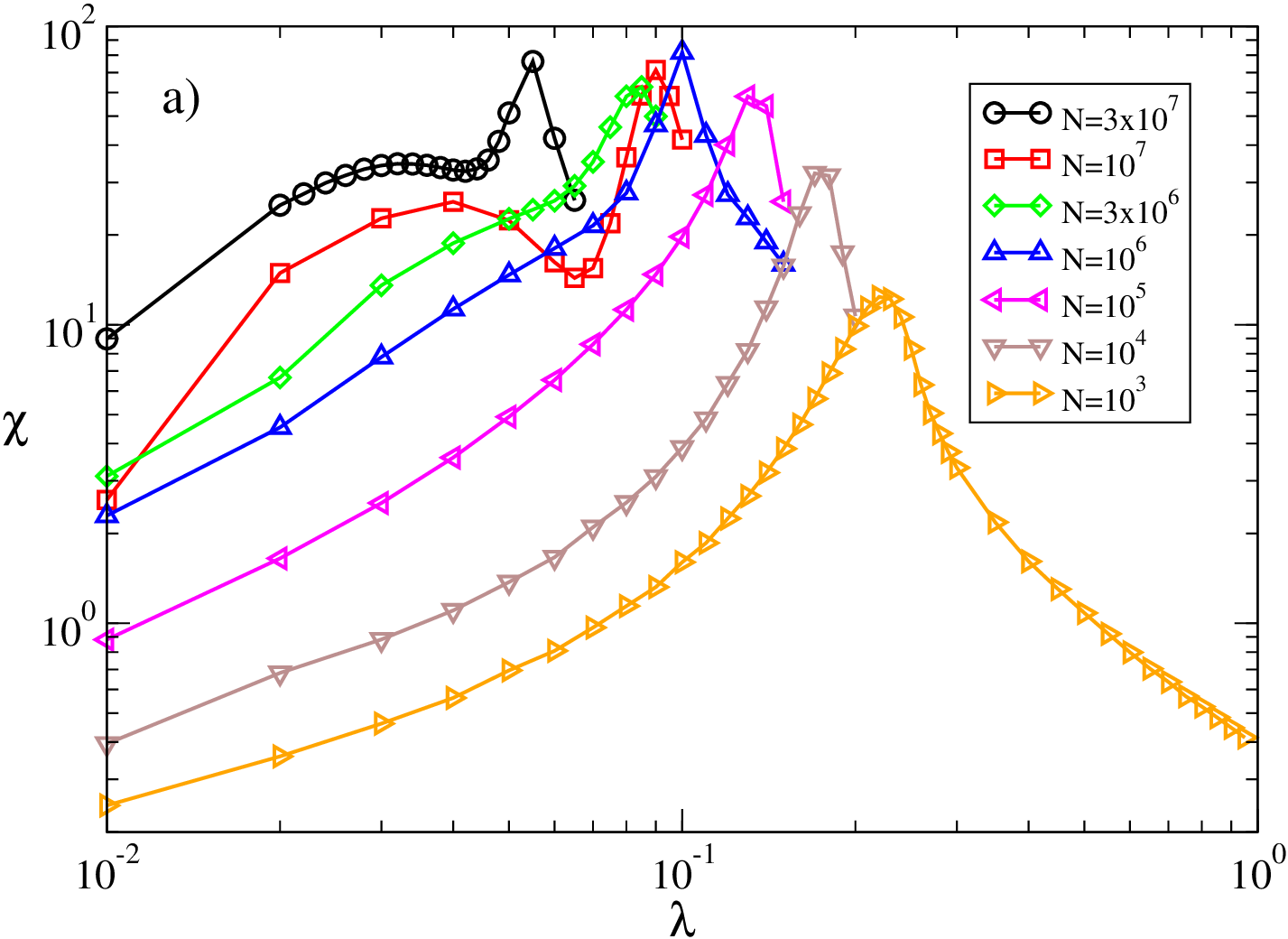}
\\~\\
\includegraphics[width=8.2cm,clip=true]{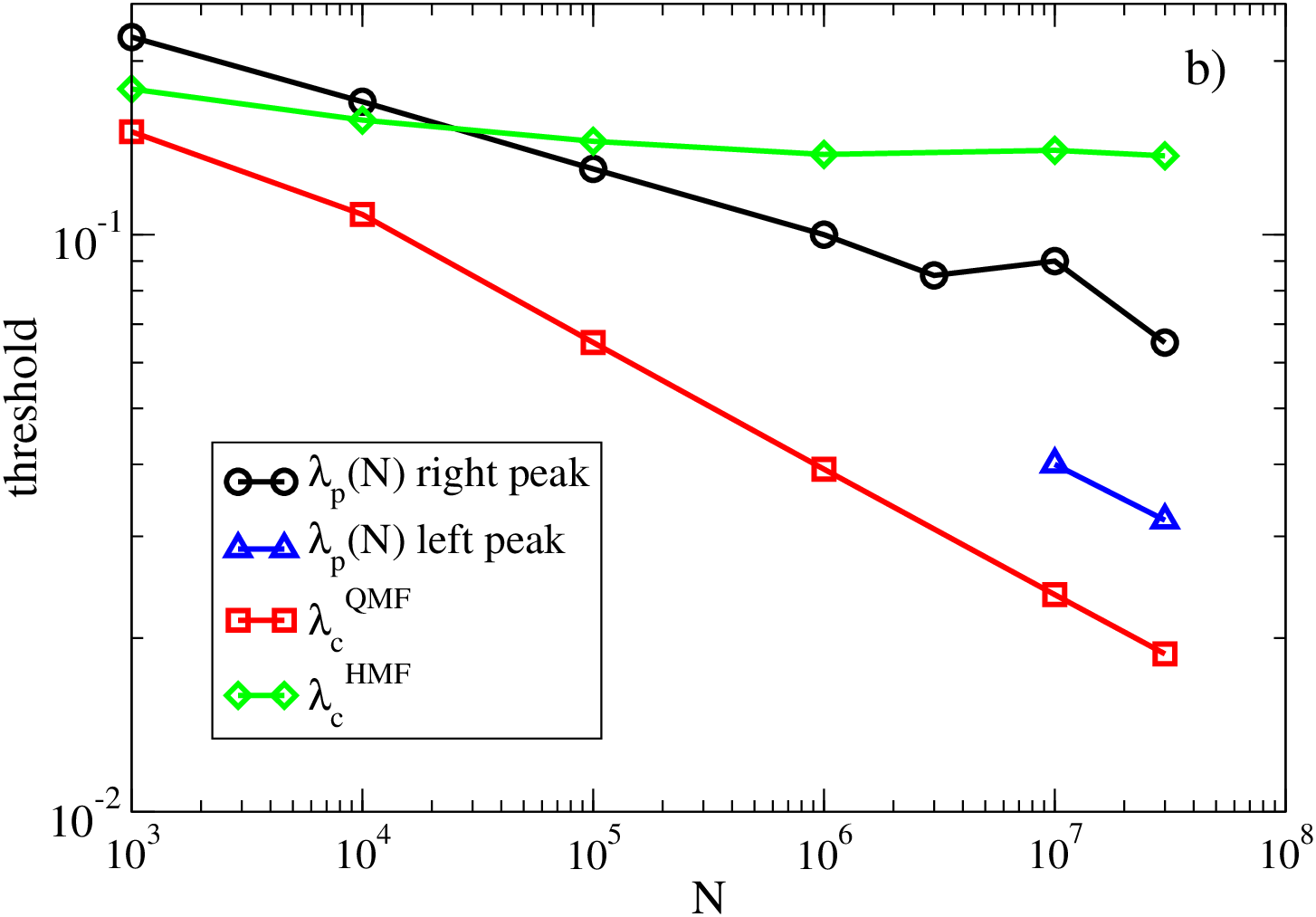}

\caption{(Color online) (a) Susceptibility $\chi$ as a function of
$\lambda$ for uncorrelated networks with $\gamma=3.5$ and different sizes.
(b) Effective thresholds $\lambda_p(N)$ determined by the rightmost and
leftmost (when present) susceptibility peak as a function of network
size $N$ for uncorrelated networks with $\gamma=3.5$,
compared with the HMF and QMF predictions.}
 \label{chi_g3.50}
\end{figure}

In Fig.~\ref{chi_g3.50}(a) we plot the susceptibility as a function of
$\lambda$ in networks with $\gamma=3.5$. The behavior of the
susceptibility in this regime is remarkably different from the one
observed in the case $\gamma<3$. As we can see, while for small
network sizes a well defined and unique peak is present for relatively
large values of $\lambda$, at a position quite compatible with the
prediction of HMF, 
as $N$ grows another feature emerges for smaller values of
$\lambda$, giving rise to a secondary peak for the largest considered sizes.

\begin{figure}[t]
 \centering
 \includegraphics[width=8.2cm,clip=true]{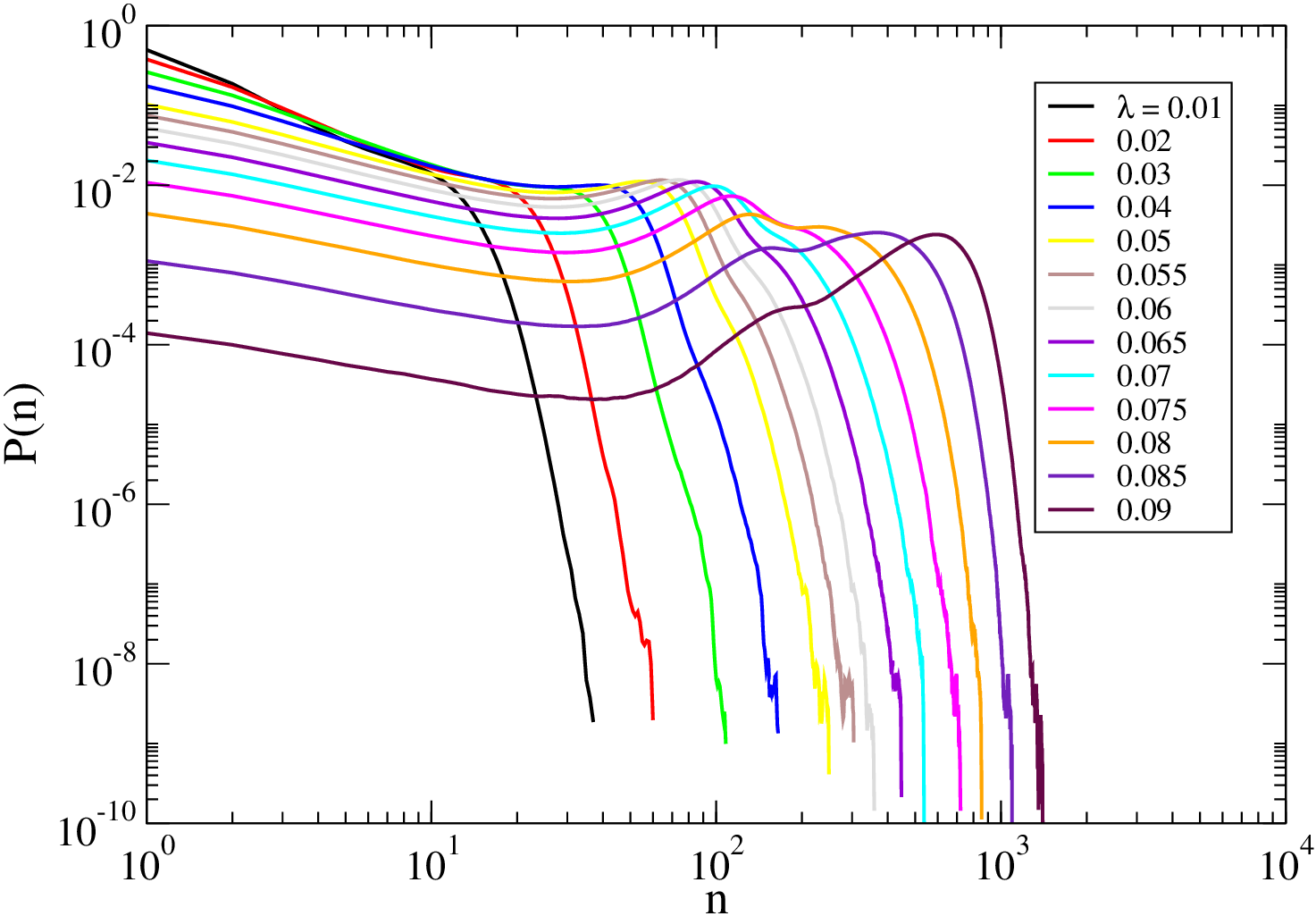}
 \caption{(Color online) Quasi-stationary distribution of the number
   of active nodes $P_n$ for an uncorrelated network of size $N=3
   \times 10^6$ and $\gamma=3.5$. Different curves are for decreasing
   values of $\lambda$ (bottom to top).}
 \label{Pofn}
\end{figure}
The evidence presented in Fig.~\ref{chi_g3.50}(a) can be better understood
with the help of the QS distribution
$\bar{P}_n$ of the order parameter (number of active nodes $n=\rho_s N$)
depicted in Fig.~\ref{Pofn}.  For large
values of $\lambda$, the distribution has a single peak (apart from
the one in $n=1$.).  As $\lambda$ is reduced, a secondary peak starts
to appear at smaller values of $n$ and rapidly takes over. Further
decreasing $\lambda$ leads to the disappearance of the all peaks
for finite $n$, which signals the transition into the absorbing state.

Figs.~\ref{chi_g3.50}(a) and~\ref{Pofn} reflect the existence of two
competing thresholds, associated to two different mechanisms
triggering the transition~\cite{Castellano12}.  The secondary peak for
small $\lambda$, whose position scales with network size as predicted
by QMF formula, see Fig.~\ref{chi_g3.50}(b), is associated to the
presence of the star-subgraph centered around the largest hub, which
for $\lambda \gtrsim 1/\Lambda_N$ is able to sustain alone the active
state \cite{Castellano10,Castellano12}.  This kind of transition
starts from a localized region~\cite{Goltsev12} and then propagates
the infection to the rest of the network. 
Its position is set by $\km$ and does not change
depending on the quenched network realization.  Notice also the
rounded shape of the susceptibility peak, reminiscent of the what is
found for star graphs (see Fig.~\ref{stargraph}). 
The peak for large
$\lambda$, which occurs not far the value predicted by HMF and is much
narrower, is associated to the set of most densely connected nodes in
the network (maximum $k$-core) collectively turning into the active
state.  The location of this transition fluctuates a lot depending on
the realization.

It is clear that for asymptotically large $N$ the first mechanism dominates. In
this limit one expects the picture to be analogous to the case $5/2<\gamma<3$
presented above: a single peak moving toward zero as $N$ increases, as predicted
by Eq.~(\ref{QMF}) (but with a different prefactor). However, the crossover to
this stage is very slow and values of $N$ much larger than those attainable with
our computational resources would be needed to check in detail the accuracy of
Eq.~(\ref{QMF}) in this regime.

\section{Conclusions}

In this paper we have presented a large scale numerical analysis of the SIS
model on networks. Our approach presents two improvements over previous
numerical studies of the SIS. Firstly, we have implemented the quasi-stationary 
state (QS) method, which allows to overcome the problems associated to simulations
based on surviving averages, yielding far better statistics with much smaller
uncertainties. Secondly, to overcome the problems that traditional finite-size
scaling analysis face in front of a vanishing critical point, we have instead
considered the susceptibility $\chi$, whose maximum value provides a numerical
estimate $\lambda_p$ of the threshold.  The combination of the QS and
susceptibility peak methods leads to numerically accurate threshold estimates,
as we have checked in the case of annealed networks, in which the exact value of
the SIS threshold is known. The accuracy of our results allows us to discuss in
detail the relative performance of two candidate theoretical solutions to the
SIS model on networks, namely the heterogeneous mean-field (HMF) and the
quenched mean-field (QMF) theories.

By considering the very simple case of the random regular graph, our analysis
shows that even for homogeneous networks, which have a finite threshold for
large sizes, both HMF and QMF theories may provide inaccurate results. This
occurs because dynamical correlations play a role in determining the threshold
value, but they are disregarded by the theoretical approaches.

In the case of strongly heterogeneous networks (the star graph), QMF theory is
sufficient to yield the correct scaling, but errs in the associated prefactor,
again due to dynamical correlations.

Turning to the more interesting case of random uncorrelated scale-free networks,
our numerical results indicate that both HMF and QMF provide asymptotically
exact expressions for the epidemic threshold in the case $\gamma<5/2$.
In the region $5/2 < \gamma < 3$, both theories are quite close for
the investigated sizes but QMF is able to
reproduce the scaling of the threshold with network size, erring only
in the numerical
prefactor. The analysis of the more complex case $\gamma>3$ leads to a
picture in agreement with the presence of two epidemic activating mechanisms
discussed in Ref.~\cite{Castellano12}. Here, the susceptibility presents for
small network sizes a peak at large $\lambda$, close to the HMF
prediction, which tends to a finite limit but largely fluctuates from sample to
sample. This peak is associated to the activation of the epidemics in the
network by the set of most densely connected nodes in the network (maximum
$k$-core).
For large sizes, a secondary incipient peak appears at small $\lambda$,
which is described by QMF
and asymptotically overcomes the other peak at sufficiently large $N$. 
The secondary peak is associated to the epidemic
activation from the most connected node in the network which, as center of a
star graph of size $\km+1$, is able, all alone, to sustain activity in the whole
network.

From our numerical assessment of the validity of HMF and QMF
theories we can conclude that, while QMF represents a notable
improvement over the 
HMF approach, it is still unable to yield a precise determination of
thresholds, apart from the special case $\gamma<5/2$. This calls thus
for improved analytical approximations, which should include in an
explicit manner the effects of dynamical correlations between
neighboring nodes. Progress has been done recently along this
path~\cite{Eames01102002,Gleeson11,Cator12}, but these methods are
easily applicable only to small networks, so that the precise
theoretical determination of the SIS epidemic threshold for large
networks remains essentially an open problem.

\begin{acknowledgments}
  SCF thanks financial support of the Brazilian agencies CNPq and
  FAPEMIG.  
  CC acknowledges financial support from the European Science Foundation
  under project DRUST.
  R.P.-S.  acknowledges financial support from the Spanish
  MEC, under project No. FIS2010-21781-C02-01; the Junta de
  Andaluc\'{\i}a, under project No. P09-FQM4682; and additional
  support through ICREA Academia, funded by the Generalitat de
  Catalunya.
\end{acknowledgments}

\bibliographystyle{apsrev4-1}
%\bibliography{sis}

%Merlin.mbs v4.21 2009-07-09.
%

\end{document}